\documentclass[prb,onecolumn,showpacs,preprintnumbers,amsmath,amssymb,floatfix]{revtex4}

\usepackage{graphicx}

\begin{document}

\title{Quantum Monte Carlo, Density Functional Theory, and
Pair Potential Studies of Solid Neon}

\author{N.~D.~Drummond and R.~J.~Needs}

\affiliation{TCM Group, Cavendish Laboratory, University of Cambridge,
J.~J.~Thomson Avenue, Cambridge CB3 0HE, United Kingdom}

\date{\today}

\begin{abstract}

We report quantum Monte Carlo (QMC), plane-wave density-functional
theory (DFT), and interatomic pair-potential calculations of the
zero-temperature equation of state (EOS) of solid neon.  We find that
the DFT EOS depends strongly on the choice of exchange-correlation
functional, whereas the QMC EOS is extremely close to both the
experimental EOS and the EOS obtained using the best semiempirical
pair potential in the literature.  This suggests that QMC is able to
give an accurate treatment of van der Waals forces in real materials,
unlike DFT\@.  We calculate the QMC EOS up to very high densities,
beyond the range of values for which experimental data are currently
available.  At high densities the QMC EOS is more accurate than the
pair-potential EOS\@.  We generate a different pair potential for neon
by a direct evaluation of the QMC energy as a function of the
separation of an isolated pair of neon atoms.  The resulting pair
potential reproduces the EOS more accurately than the equivalent
potential generated using the coupled-cluster CCSD(T) method.

\end{abstract}


\maketitle

\section{Introduction}

One of the most important goals of \textit{ab initio} computational
electronic-structure theory is the development of accurate methods for
describing interatomic bonding.  Quantum Monte Carlo (QMC) techniques
are useful in this regard as they can provide a highly accurate
description of electron correlation effects.  Although QMC methods are
computationally expensive, they can be applied to systems which are
large enough to model condensed matter.

In this study we have considered solid neon, in which the bonding
arises from the competition between short-range repulsion and the van
der Waals attraction between the atoms.  High-quality experimental
measurements of the equation of state (EOS) exist, which can be used
as reference data.  Furthermore, various neon pair potentials have
been developed using experimental and theoretical data, which can also
be used for comparison purposes.  Solid neon is therefore an ideal
system in which to test the descriptions of van der Waals bonding and
short-range repulsion offered by various theoretical methods.

We have calculated theoretical EOS's for crystalline neon using the
QMC and density-functional theory (DFT) \textit{ab initio}
electronic-structure methods as well as various interatomic pair
potentials.  Standard DFT methods do not describe van der Waals
bonding accurately, but they might be expected to work quite well at
high densities, where the short-range repulsion dominates.  The
high-pressure properties of neon are of some experimental interest,
because neon is often used as a pressure-conducting medium in
diamond-anvil-cell experiments.\cite{neon_dac_exps}  We have therefore
extended the range of our QMC EOS for neon to very high pressures
(about 400~GPa).

The zero-point energy (ZPE) of the lattice-vibration modes makes a
small but important contribution to the total energy of solid neon.
We have therefore studied the lattice dynamics of solid neon within
the quasiharmonic-phonon approximation and within the Einstein
approximation, using DFT methods and pair potentials.

For some time there has been considerable interest in developing neon
pair potentials in order to test the accuracy of theoretical methods
for calculating the properties of materials.\cite{rgs} We have
performed a direct calculation of the neon pair potential using QMC\@.
We compare the accuracy of the EOS predicted by this pair potential
with the results obtained using other pair potentials, including one
obtained from coupled-cluster CCSD(T) calculations.\cite{cybulski}

Detailed information about our computational methodologies is given in
Sec.~\ref{sec:methodology} and DFT calculations of the phase stability
and band gap of solid neon are reported in Sec.~\ref{sec:prelims}.
The calculation of a neon pair potential using QMC is described in
Sec.~\ref{sec:dmc_pair_pot}.  The lattice dynamics of solid neon are
studied in Sec.~\ref{sec:neon_latt_dyn}.  We compare the EOS's
obtained using different methods in Sec.~\ref{sec:neon_EoS}.  Finally,
we draw our conclusions in Sec.~\ref{sec:conclusions}.

Hartree atomic units (a.u.)\ are used throughout, in which the Dirac
constant, the magnitude of the electronic charge, the electronic mass,
and $4\pi$ times the permittivity of free space are unity: $\hbar =
|e| = m_e = 4 \pi \epsilon_0 = 1$.

\section{Methodology \label{sec:methodology}}

\subsection{DFT calculations}

\subsubsection{DFT total-energy calculations}

Our DFT calculations were performed using the \textsc{castep}
plane-wave-basis code.\cite{newcastep} The local-density approximation
(LDA) and Perdew-Burke-Ernzerhof (PBE)
generalized-gradient-approximation\cite{pbe} exchange-correlation
functionals were used.  The Ne$^{8+}$ ionic cores were represented by
ultrasoft pseudopotentials.\cite{newcastep} The EOS calculations were
performed using a $4 \times 4 \times 4$ Monkhorst-Pack ${\bf k}$-point
mesh and a plane-wave cutoff energy of 200~a.u., for which the DFT
energies have converged to about 7 significant figures.  The
self-consistent-field calculations were judged to have converged when
the fractional change in the energy was less than $10^{-11}$.  The DFT
band-gap calculations reported in Sec.\ \ref{sec:bandgap} were
performed using the same parameters, except that the plane-wave cutoff
energies ranged from 100~a.u.\ for the lowest densities to 800~a.u.\
for the highest densities.

\subsubsection{DFT force-constant calculations
  \label{sec:dft_fc_calcs}}

We used the quasiharmonic approximation\cite{wallace} to evaluate the
DFT ZPE of the lattice-vibration modes and we used the method of
finite displacements and the Hellmann-Feynman theorem to evaluate the
density-dependent force constants.  Symmetry and Newton's third law
were imposed iteratively on the matrix of force
constants.\cite{ackland} The DFT force-constant calculations were
carried out using a plane-wave cutoff energy of 60~a.u., a
$3\times3\times3$ Monkhorst-Pack ${\bf k}$-point mesh, and ultrasoft
pseudopotentials.\cite{newcastep} The force constants were converged
to about $10^{-6}$~a.u.\ with respect to the plane-wave cutoff energy
and the ${\bf k}$-point mesh.  In the production force-constant
calculations, the displacement of the neon atom from its equilibrium
position was $2.12$\% of the nearest-neighbor distance in each case,
which ensures that anharmonic effects are negligible.  The
force-constant calculations were carried out in both $2 \times 2
\times 2$ and $3 \times 3 \times 3$ supercells of the primitive unit
cell, and the difference in the resulting ZPE's was found to be
negligible.  The dispersion curves shown in
Sec.~\ref{sec:neon_latt_dyn} were produced using a $3\times 3 \times
3$ supercell, while the ZPE's that were combined with the
static-lattice EOS's were calculated in a $2 \times 2 \times 2$
supercell.

\subsubsection{DFT orbital-generation calculations}

DFT-LDA calculations were performed in order to generate orbitals for
the trial wave functions used in the QMC calculations.  The QMC
calculations made use of relativistic Hartree-Fock neon
pseudopotentials,\cite{jrt_psps_1,jrt_psps_2} and these were also used
in the DFT orbital-generation calculations.  The Hartree-Fock
pseudopotentials are much harder than the ultrasoft pseudopotentials.
Plane-wave cutoffs in excess of 250~a.u.\ were used in each
orbital-generation calculation, so the DFT energy was converged to
around $10^{-3}$~a.u.  This cutoff is very large by the normal
standards of DFT calculations, but there is evidence that using large
basis sets reduces the variance of the energy in QMC
calculations.\cite{alfe_2005}

\subsection{QMC calculations}

\subsubsection{VMC and DMC methods}

In the variational quantum Monte Carlo (VMC) method, expectation
values are calculated using an approximate trial wave function, the
integrals being performed by a Monte Carlo technique.  In diffusion
quantum Monte Carlo\cite{ceperley_1980,foulkes} (DMC) the
imaginary-time Schr\"odinger equation is used to evolve an ensemble of
electronic configurations towards the ground state.  The fermionic
symmetry is maintained by the fixed-node
approximation,\cite{anderson_1976} in which the nodal surface of the
wave function is constrained to equal that of a trial wave function.
Furthermore, the use of nonlocal pseudopotentials to represent the
Ne$^{8+}$ cores necessitates the use of the locality
approximation,\cite{hurley_1987} which leads to errors that are second
order in the quality of the trial wave function.\cite{mitas}

Our QMC calculations were performed using the \textsc{casino}
code.\cite{casino} The trial wave functions were of Slater-Jastrow
form, with the orbitals in the Slater wave function being taken from
DFT calculations and the free parameters in the Jastrow factor being
optimized by minimizing the unreweighted variance of the
energy.\cite{umrigar_1988a,ndd_newopt}  The DFT-generated orbitals
were represented numerically using splines on a grid in real space
rather than an expansion in plane waves in order to improve the
scaling of the QMC calculations with system
size.\cite{alfe_blips,williamson_lin_scaling}  The Jastrow factors
consisted of isotropic electron-electron, electron-nucleus, and
electron-electron-nucleus terms.\cite{ndd_jastrow}  The
electron-electron terms describe long-ranged correlations and
therefore play the most important role in describing van der Waals
forces.

\subsubsection{Finite-size bias \label{sec:finite_size_bias}}

The QMC simulations of crystalline neon were carried out in supercells
of finite size subject to periodic boundary conditions.  The
electrostatic energy of each electron configuration was calculated
using the Ewald method.\cite{ewald} The QMC energy per atom obtained
in a finite cell differs from the energy per atom of the infinite
crystal due to \textit{single-particle} finite-size effects and
\textit{Coulomb} finite-size effects.  The former result from the fact
that the allowed ${\bf k}$ points for the Bloch orbitals form a
discrete lattice, so that the single-particle energy components change
when the size of the simulation supercell is changed.  The latter,
which are the more important in insulators, are caused by the
interaction of the charged particles with their periodic images.  At
any given instant, each electron feels itself to be part of an
infinite crystal of electrons.\cite{ceperley_1978,kent_1999} The
resulting bias is negative, and is generally believed to fall off as
$1/N$, where $N$ is the number of atoms in the simulation
cell.\cite{ceperley_1980,zong_2002,ceperley_1987,rajagopal_1995}

In order to eliminate the finite-size bias, simulations were carried
out in supercells consisting of $3 \times 3 \times 3$ and $4 \times 4
\times 4$ primitive unit cells.  The error in the DFT results arising
from the use of a $3 \times 3 \times 3$ ${\bf k}$-point mesh is small
(about $0.0001$~a.u.), so we conclude that single-particle finite-size
effects are negligible.  The assumed form of the Coulomb finite-size
bias was therefore used to extrapolate the results to infinite system
size.  The static-lattice energy per atom in the infinite-system limit
is given by
\begin{equation}
E_\infty^{\rm SL}(V)=E_N^{\rm SL}(V)+\frac{b(V)}{N},
\label{eqn:energy_extrap}
\end{equation}
where $E_N^{\rm SL}(V)$ is the Vinet fit (see
Sec.~\ref{sec:EOS_models}) to the DMC static-lattice energy-volume
data obtained in a set of $N$-atom simulation supercells, $V$ is the
primitive-cell volume, and $b(V)$ is a parameter determined by
fitting.  Since we only have energy-volume data for two different
system sizes, $N$ and $M$, we may eliminate $b(V)$ and write
\begin{equation}
E_\infty^{\rm SL}(V)=\frac{NE_N^{\rm SL}(V)-ME_M^{\rm SL}(V)}{N-M}.
\label{eqn:energy_extrap_practice}
\end{equation}
The pressure due to the static-lattice energy at infinite system size
is given by
\begin{equation}
p_\infty^{\rm SL}(V) \equiv -\frac{dE_\infty^{\rm SL}}{dV} =
\frac{Np_N^{\rm SL}(V)-Mp_M^{\rm SL}(V)}{N-M},
\label{eqn:pressure_extrap}
\end{equation}
where $p_N^{\rm SL}(V) \equiv -dE_N^{\rm SL}/dV$ is the static-lattice
pressure in an $N$-atom simulation supercell.

\begin{figure}
\begin{center}
\includegraphics*{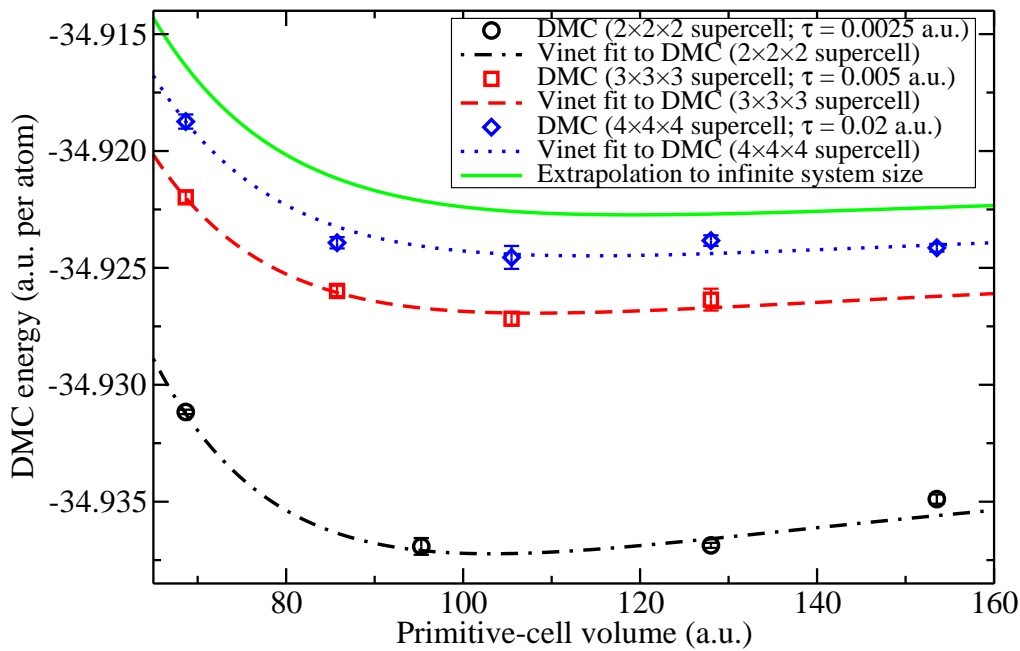}
\caption{(Color online) Low-density static-lattice DMC energy of FCC
neon as a function of volume, evaluated in simulation supercells
consisting of $n \times n \times n$ primitive unit cells using
different time steps $\tau$.
\label{fig:neon_e_of_V}}
\end{center}
\end{figure}

\begin{figure}
\begin{center}
\includegraphics*{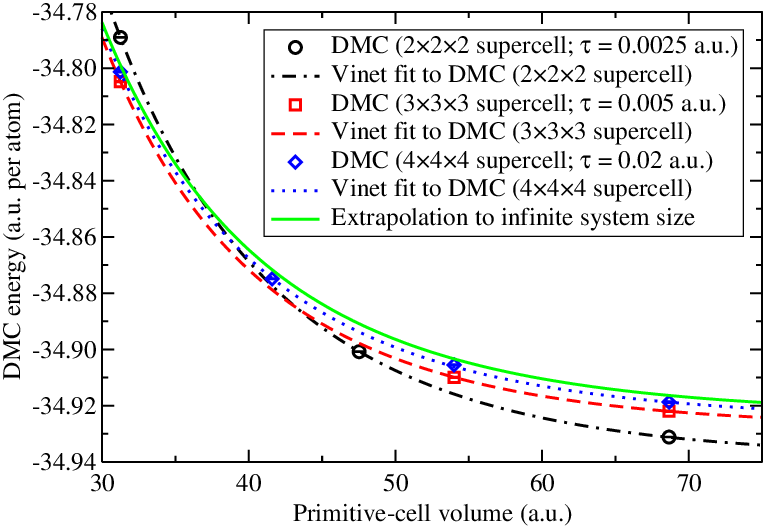}
\caption{(Color online) Same as Fig.~\ref{fig:neon_e_of_V}, but at
  higher densities.
\label{fig:neon_e_of_V_highden}}
\end{center}
\end{figure}

\begin{figure}
\begin{center}
\includegraphics*{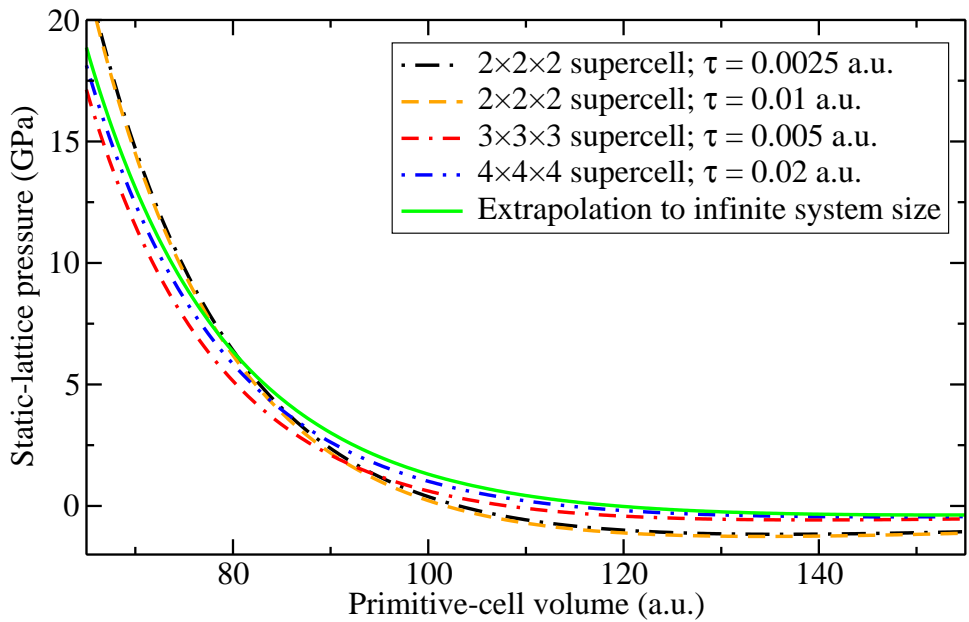}
\caption{(Color online) Low-density static-lattice pressure obtained
by fitting Vinet EOS's to the DMC energies obtained in simulation
supercells consisting of $n \times n \times n$ primitive unit cells
and at different time steps $\tau$.  The pressure extrapolated to
infinite system size is also shown.
\label{fig:neon_qmc_pressures}}
\end{center}
\end{figure}

\begin{figure}
\begin{center}
\includegraphics*{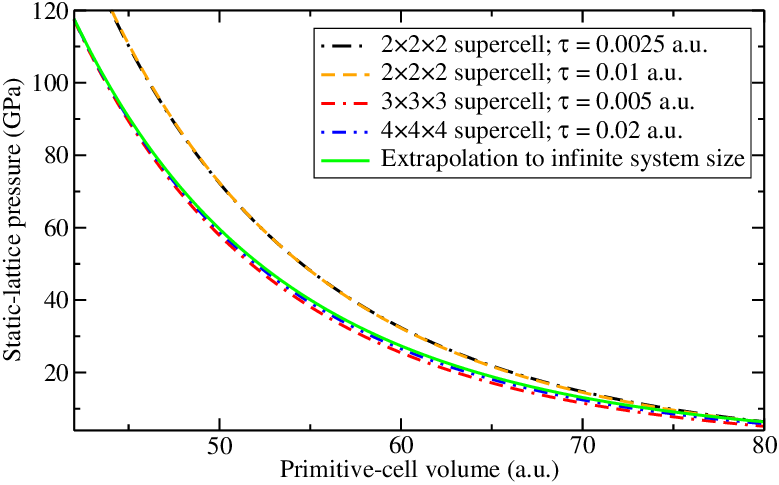}
\caption{(Color online) Same as Fig.~\ref{fig:neon_qmc_pressures}, but
  at higher densities.  \label{fig:neon_qmc_pressures_highden}}
\end{center}
\end{figure}

The zero-temperature, static-lattice energy-volume curves of neon,
calculated using DMC in different sizes of simulation supercell are
shown in Figs.~\ref{fig:neon_e_of_V} and
\ref{fig:neon_e_of_V_highden}. Vinet EOS's are fitted to the data.
The corresponding pressure-volume data are shown in
Figs.~\ref{fig:neon_qmc_pressures} and
\ref{fig:neon_qmc_pressures_highden}.  It can be seen that the DMC
pressure-volume curves converge steadily with system size, and that
the DMC pressure extrapolated to infinite system size using
Eq.~(\ref{eqn:pressure_extrap}) is close to the pressure of the $4
\times 4 \times 4$ supercell.  This implies that the error introduced
by the extrapolation is small, because the extrapolation is itself a
small correction.

\subsubsection{Time-step bias}

\begin{figure}
\begin{center}
\includegraphics*{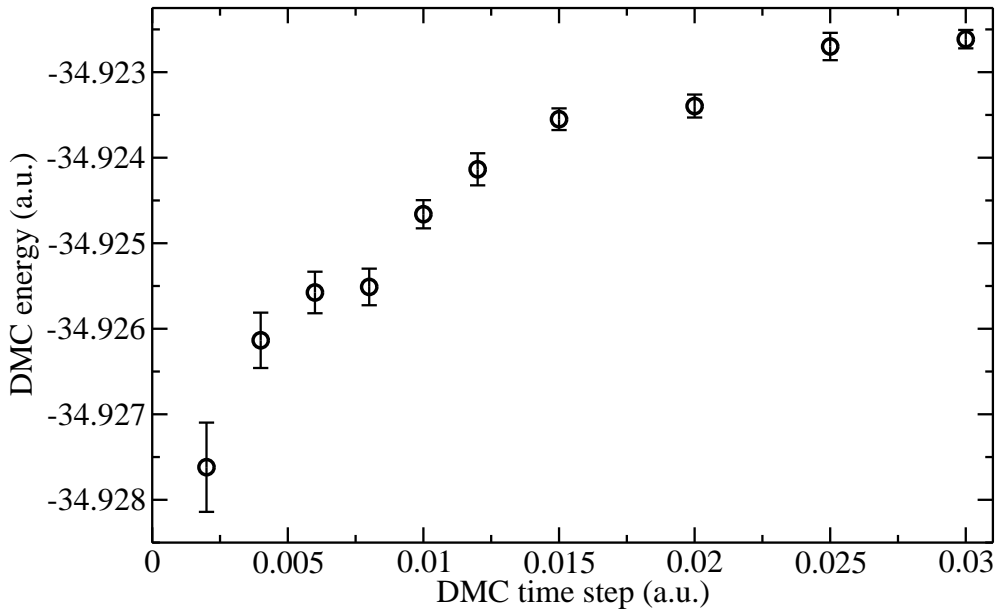}
\caption{DMC energy of solid neon plotted against time step for a
simulation supercell consisting of $3 \times 3 \times 3$ FCC primitive
unit cells.  The primitive-cell volume is $85.75$~a.u.  960
configurations were used in the DMC simulations.
\label{fig:neon333_dt_bias}}
\end{center}
\end{figure}

The fixed-node DMC Green's function is only exact in the limit of zero
time step; the use of a nonzero time step biases the DMC energy.  An
example of the bias in the DMC energy of a pseudoneon crystal is shown
in Fig.~\ref{fig:neon333_dt_bias}.  On the other hand, as shown by the
$2 \times 2 \times 2$ supercell results in
Figs.~\ref{fig:neon_qmc_pressures} and
\ref{fig:neon_qmc_pressures_highden}, the pressure is very insensitive
to the time step.  We used time steps of $0.005$~a.u.\ and
$0.02$~a.u.\ in our production calculations for the $3 \times 3 \times
3$ and $4 \times 4 \times 4$ supercells, respectively.  A time-step of
$0.002$~a.u.\ was used in the DMC pair-potential calculations.  The
target population was at least 320 configurations in each case, while
a target population of 1000 configurations was used for the
pair-potential-generation calculations.

\subsection{Pair-potential calculations}

\subsubsection{Forms of neon pair potential}

We have used the following forms of pair potential: (i) the HFD-B
potential proposed by Aziz and Chen\cite{aziz_1977} with the parameter
values given by Aziz and Slaman\cite{aziz_1989,footnote_aziz}; (ii)
the form of potential proposed for helium by Korona \textit{et
al.},\cite{korona} containing the parameter values determined by
Cybulski and Toczy{\l}owski\cite{cybulski} using all-electron
double-excitation coupled-cluster theory with a non-iterative
perturbational treatment of triple excitations (CCSD(T)) and an
av5z$+$ Gaussian basis set; and (iii) a fit of the potential of Korona
\textit{et al.}\cite{korona}\ to our DMC energy data, as described in
Sec.~\ref{sec:dmc_pair_pot}.  We believe the HFD-B pair potential to
be the most accurate neon pair potential in the literature to date.

\subsubsection{Static-lattice energy-volume curve using pair potentials}

Let the pair potential between two neon atoms at ${\bf R}$ and ${\bf
R}^\prime$ be $\phi(|{\bf R}-{\bf R}^\prime|)$.  Let $A$ be a large
radius.  We evaluate the static-lattice energy per atom as
\begin{equation}
E^{\rm SL} \approx \frac{1}{2} \left( \sum_{0<|{\bf R}|<A} \phi(|{\bf
R}|) + \frac{N_{\rm Ne}}{V} \int_A^\infty 4 \pi r^2 \phi(r) \, dr
\right),
\label{eqn:ppsl}
\end{equation}
where the $\{ {\bf R} \}$ are the lattice sites and $N_{\rm Ne}/V$ is
the number density of neon atoms.  This expression becomes exact as
$A$ goes to infinity.  The integral in Eq.~(\ref{eqn:ppsl}) was
evaluated analytically for each pair potential, while the sum was
evaluated by brute force.  $A$ was increased until $E^{\rm SL}$
converged.

\subsubsection{Force-constant calculations using pair potentials}

Pair potentials were used to generate force-constant data for a
quasiharmonic\cite{wallace} calculation of the ZPE of neon.  The
method of finite displacements\cite{ackland} was used to generate the
force constants in a finite supercell subject to periodic boundary
conditions.  It was ensured that the force constants were highly
converged with respect to the size of the displacements and the number
of periodic images of the neon atoms that contributed to the force
constants.  Following the evaluation of the force constants, the
calculation of the ZPE proceeded as described in
Sec.~\ref{sec:dft_fc_calcs}.

\subsection{EOS models \label{sec:EOS_models}}

Let $E(V)$ be the total energy of a neon crystal as a function of
primitive-cell volume $V$.  It has previously been
noticed\cite{tsuchiya} that a Vinet EOS of the form
\begin{equation}
E(V)= - \frac{4B_0 V_0}{(B_0^\prime -1)^2} \left( 1-\frac{3}{2}
(B_0^\prime-1) \left(1-\left(\frac{V}{V_0} \right)^{1/3} \right)
\right) \exp \left( \frac{3}{2} (B_0^\prime -1) \left(1-\left(
\frac{V}{V_0} \right)^{1/3} \right) \right) + C,
\label{eqn:Vinet_EOS}
\end{equation}
where the zero-pressure volume $V_0$, bulk modulus $B_0$,
pressure-derivative of the bulk modulus $B_0^\prime$, and integration
constant $C$ are fitting parameters, gives a better fit than a
third-order Birch-Murnaghan EOS,
\begin{equation}
E(V)= - \frac{9}{16} B_0 \left( (4-B_0^\prime) \frac{V_0^3}{V^2} -
(14-3B_0^\prime) \frac{V_0^{7/3}}{V^{4/3}}+(16-3B_0^\prime)
\frac{V_0^{5/3}}{V^{2/3}} \right) + C,
\label{eqn:BM_EOS}
\end{equation}
to DFT results for solid neon.  In some cases the Vinet EOS gives a
lower $\chi^2$ value when fitted to our DMC data than the
Birch-Murnaghan EOS; in others it gives a higher $\chi^2$ value.  For
example, using DMC data obtained in simulation cells consisting of $2
\times 2 \times 2$ primitive cells and a time step of $0.01$~a.u., the
Vinet and Birch-Murnaghan EOS models give $\chi^2$ values of $10.0914$
and $35.3561$, respectively, whereas at a time step of $0.0025$~a.u.\
the EOS models give $\chi^2$ values of $20.6609$ and $4.8967$,
respectively.  The resulting pressure-volume curves are essentially
indistinguishable in each case, however.  To be consistent, we have
fitted Vinet EOS's to all of our theoretical data.

\section{DFT study of phase stability and band gap \label{sec:prelims}}

\subsection{Phase transitions in solid neon \label{sec:phase_transitions}}

We have compared the DFT energies of face-centered cubic (FCC) and
hexagonal close-packed (HCP) phases of solid neon.  For HCP neon the
lattice-parameter ratio $c/a$ was optimized, but the optimal ratio
always turned out to be $\sqrt{8/3}$, which is the ratio appropriate
for an ideal HCP lattice.  The DFT energy difference between the FCC
and HCP phases is typically less than $0.0005$~a.u.:\ too small for us
reliably to identify any phase transition.  Experimentally, Hemley
\textit{et al.}\cite{hemley_1989}\ have found that solid neon adopts
the FCC phase up to pressures of at least 110~GPa at 300~K\@.  We have
therefore used the FCC lattice in all of our calculations, apart from
those described in this section.

\subsection{Band gap of solid neon \label{sec:bandgap}}

\begin{figure}
\begin{center}
\includegraphics*{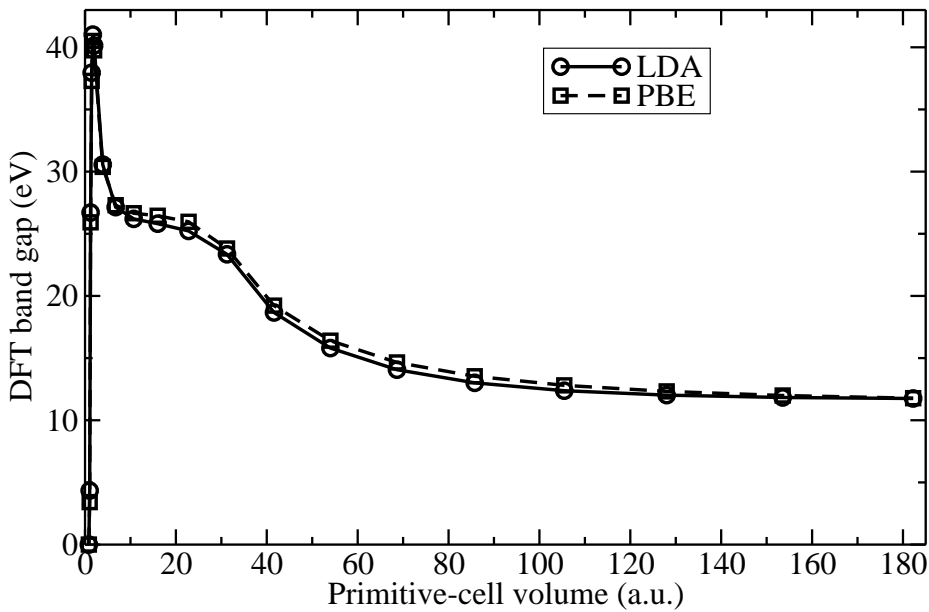}
\caption{DFT band gap of FCC neon against the primitive-cell volume,
calculated using the LDA and PBE exchange-correlation functionals.
\label{fig:bandgap_v_vol}}
\end{center}
\end{figure}

The band gap of solid neon, calculated using DFT, is shown in
Fig.~\ref{fig:bandgap_v_vol}.  The band gap is large at the
equilibrium volume, and increases significantly when the material is
compressed.  The DFT calculations predict that neon is still an
insulator when it is compressed to a primitive-cell volume of 2~a.u.,
corresponding to a pressure of about 366~TPa.  The use of the
ultrasoft neon pseudopotential (with a core radius of 0.9~a.u.)\
probably causes the DFT results to become unreliable at such high
densities; nevertheless, our results indicate that the metalization
pressure of neon is of the order of hundreds of TPa.  Hemley
\textit{et al.}\cite{hemley_1989}\ concluded that neon remains a
wide-gap insulator over the range of pressures that they studied using
diamond-anvil cells (up to 110.4~GPa), while Hawke \textit{et
al.}\cite{hawke}\ used a magnetic-flux compression device to show that
solid neon remains an insulator up to at least 500~GPa.

The DFT-LDA and DFT-PBE band gaps at the experimental equilibrium
primitive-cell volume (150~a.u.)\ are 11.85~eV and 12.04~eV,
respectively, compared with the experimentally determined value of
21.51~eV\@.\cite{runne} As usual, DFT substantially underestimates the
band gap.  The \textit{GW} band gap of 20.04~eV, calculated by
Galami\'c-Mulaomerovi\'c and Patterson, is relatively
accurate.\cite{patterson} The DMC method can also be used to perform
highly accurate band-gap calculations,\cite{williamson1998,towler2000}
although we have not done this for neon.

\section{DMC-calculated pair potential for neon \label{sec:dmc_pair_pot}}

\begin{figure}
\begin{center}
\includegraphics*{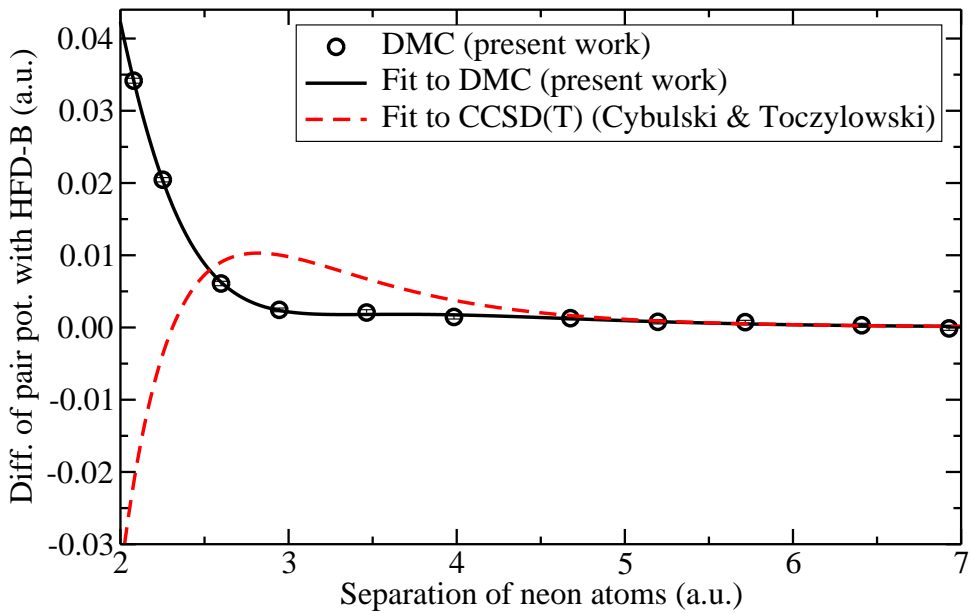}
\caption{(Color online) Neon pair potential, calculated using DMC\@.
The statistical error bars on the DMC results are smaller than the
symbols.  A fit of the form of pair potential proposed by Korona
\textit{et al.}\cite{korona}\ to the DMC data is also shown, as is the
pair potential generated by Cybulski and Toczy{\l}owski using CCSD(T)
theory.\cite{cybulski}  All are plotted relative to the HFD-B pair
potential of Aziz and Slaman.\cite{aziz_1989}
\label{fig:dmc_pair_pot}}
\end{center}
\end{figure}

The difference between the DMC pair potential, evaluated as the
fixed-nucleus total energy of a neon dimer, and the HFD-B potential is
shown in Fig.~\ref{fig:dmc_pair_pot}.  The DMC energy data have been
offset by a constant that was determined by fitting the data to a
pair-potential model.\cite{footnote_offset} We have used the form of
potential proposed by Korona \textit{et al.}\ for helium,\cite{korona}
which can be written as
\begin{equation}
\phi(r) = A \exp \left( -\alpha r + \beta r^2 \right) + \sum_{n=3}^8
f_{2n}(r,b) \frac{C_{2n}}{r^{2n}},
\end{equation}
where $r$ is the separation of the neon atoms. The dispersion
coefficients $C_{2n}$ are taken from Cybulski and
Toczy{\l}owski\cite{cybulski} ($C_6=6.28174$, $C_8=90.0503$,
$C_{10}=1679.45$, $C_{12}=4.18967 \times 10^4$, $C_{14}=1.36298 \times
10^6$, and $C_{16}=5.62906 \times 10^7$) and $f_{2n}(r,b)$ is the
damping function proposed by Tang and Toennies,\cite{tang_toennies}
\begin{equation}
f_{2n}(r,b) = 1 - \exp(-br) \sum_{k=0}^{2n} \frac{(br)^k}{k!}.
\end{equation}
$A$, $\alpha$, $\beta$, and $b$ are adjustable parameters, which were
determined by a $\chi^2$ fit to the DMC data, as was the constant
offset.  The fitted parameter values (in a.u.)\ are $A=84.956788$,
$\alpha=2.0683266$, $\beta=-0.11767673$, and $b=2.6899868$.  The
difference of the resulting pair potential with the HFD-B potential is
shown in Fig.~\ref{fig:dmc_pair_pot}, as is the corresponding curve
for the CCSD(T) data.  It can be seen that over a wide range of
separations the DMC pair potential lies closer to the HFD-B pair
potential than the CCSD(T)-generated pair potential.

\section{Lattice dynamics and zero-point energy
\label{sec:neon_latt_dyn}}

By comparing the results obtained using a Lennard-Jones potential in
the harmonic approximation with the VMC\cite{footnote_vmc} results
obtained using the same potential by Hansen,\cite{hansen} Pollock
\textit{et al.}\cite{pollock}\ have demonstrated that the harmonic
approximation is valid for solid neon at high pressures.  We consider
two methods for calculating the ZPE: (i) the ZPE of quasiharmonic
phonons can be evaluated in a supercell of several primitive cells, or
(ii) the ZPE can be computed within the Einstein approximation by
evaluating the quadratic potential felt by each atom as it is
displaced from its equilibrium position with all the other atoms held
fixed.

\begin{figure}
\begin{center}
\includegraphics*{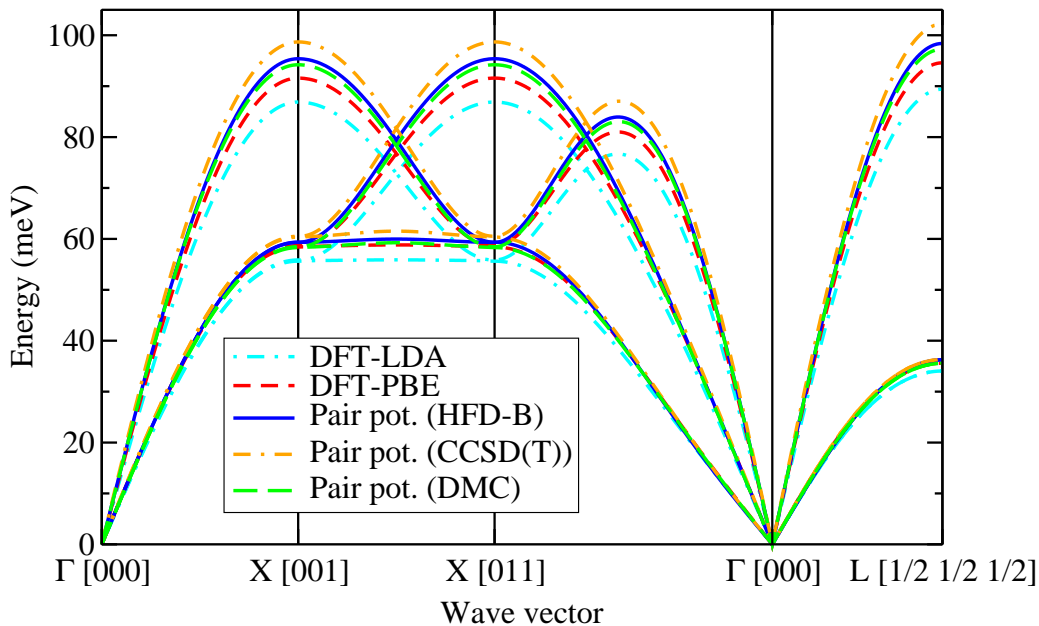}
\caption{(Color online) Phonon dispersion curves calculated using DFT
and pair potentials for FCC neon at a primitive-cell volume of
41.59375~a.u.  The Einstein frequencies evaluated using DFT-LDA,
DFT-PBE, the HFD-B pair potential, the CCSD(T) pair potential, and the
DMC pair potential are 83.10234~meV, 87.57110~meV, 63.77843~meV,
65.73911~meV, and 63.00858~meV, respectively.  The quasiharmonic ZPE
evaluated using the HFD-B potential is 0.003321155~a.u.
\label{fig:neon_disp_curve_bx2.75}}
\end{center}
\end{figure}

\begin{figure}
\begin{center}
\includegraphics*{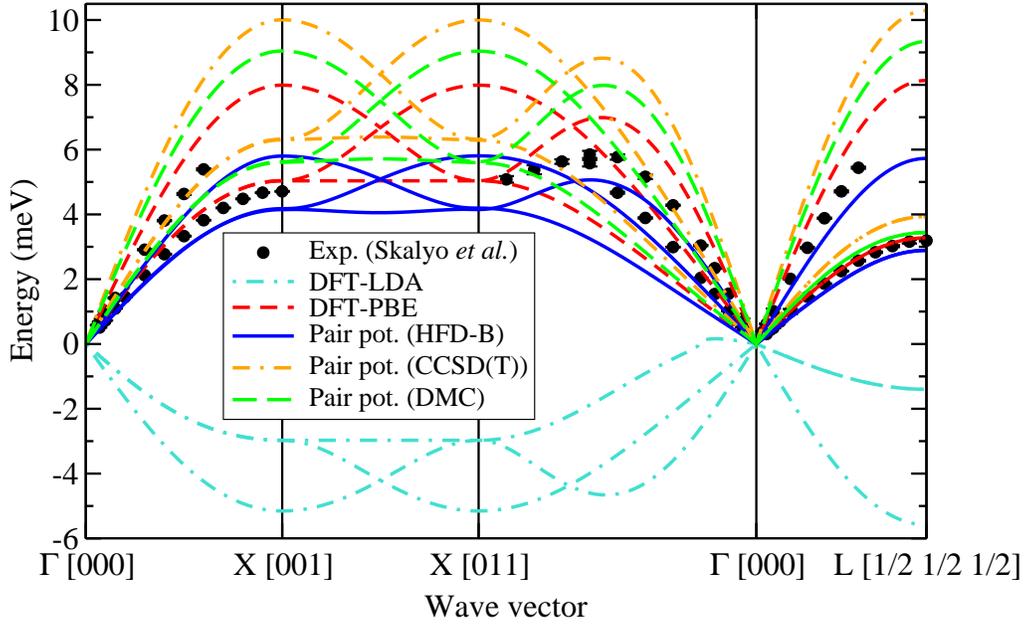}
\caption{(Color online) The same as
Fig.~\ref{fig:neon_disp_curve_bx2.75}, but with a primitive-cell
volume of 149.06894~a.u.~(close to the experimental equilibrium
density).  Experimental data from Ref.~\onlinecite{skalyo} are also
shown.  The imaginary frequencies of unstable modes are plotted as
negative numbers.  The Einstein frequencies evaluated using DFT-PBE,
the HFD-B pair potential, the CCSD(T) pair potential, and the DMC pair
potential are 5.37853~meV, 4.07895~meV, 6.73975~meV, and 6.06113~meV,
respectively.  The quasiharmonic ZPE evaluated using the HFD-B
potential is 0.000216603~a.u.
\label{fig:neon_disp_curve_equilvol}}
\end{center}
\end{figure}

Examples of phonon dispersion curves at two different densities are
shown in Figs.~\ref{fig:neon_disp_curve_bx2.75} and
\ref{fig:neon_disp_curve_equilvol}.  Inelastic neutron-scattering
data\cite{skalyo} are also shown in
Fig.~\ref{fig:neon_disp_curve_equilvol}.  At high density the DFT-LDA
and DFT-PBE dispersion curves are in good agreement, but at low
density the DFT-LDA phonon frequencies are significantly lower than
the DFT-PBE frequencies.  Unstable (imaginary) phonon modes start to
occur at a primitive-cell volume of about 133~a.u.\ in the LDA\@.  By
contrast, there are no unstable phonon modes, even at a primitive-cell
volume of 182.25~a.u., when the PBE functional is used.  The DFT and
pair-potential results are in agreement at high densities, indicating
that the DFT results are accurate in this regime.  Overall, the
DFT-PBE dispersion curves appear to be more accurate (that is, closer
to the HFD-B and experimental results) than the DFT-LDA dispersion
curves.

\begin{figure}
\begin{center}
\includegraphics*{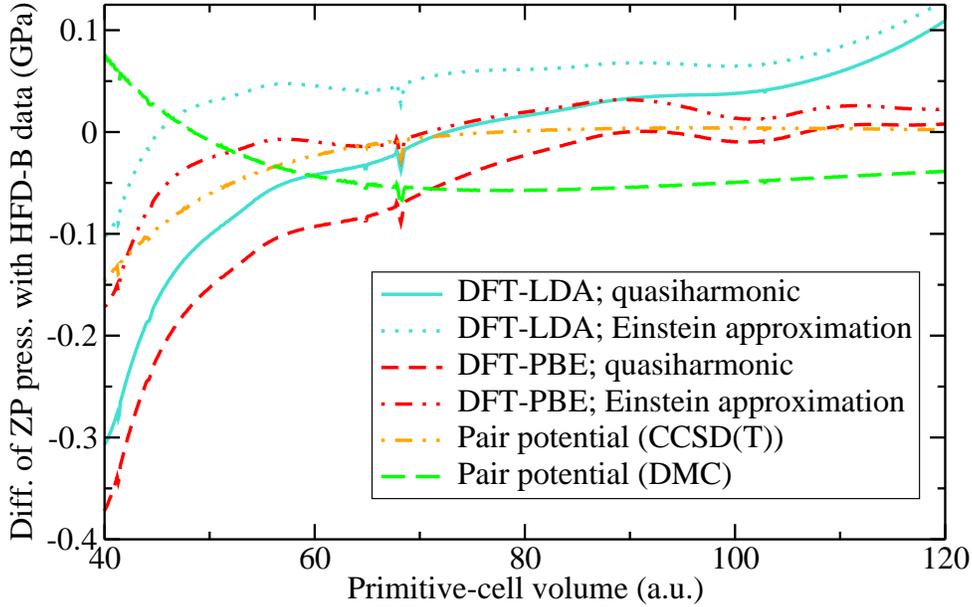}
\caption{(Color online) Difference of zero-point pressure of FCC neon
  calculated using various methods and the HFD-B result for the
  zero-point pressure.  (The noise is due to the fact that Monte Carlo
  methods were used to sample the first Brillouin zone when
  calculating the zero-point energy, and the resulting curve was
  differentiated numerically to obtain the zero-point pressure.)
\label{fig:ZPp_graph}}
\end{center}
\end{figure}

The pressure arising from the ZPE of solid neon as calculated using
different methods is plotted relative to the HFD-B results in
Fig.~\ref{fig:ZPp_graph}.  Within DFT, the Einstein approximation is
excellent.  It can be seen that the difference between the LDA and PBE
results is appreciable at low densities, but that the difference
between the Einstein and quasiharmonic zero-point pressures is more
significant at high densities.  As expected from examination of the
dispersion curves, the HFD-B zero-point pressure is closer to the
DFT-PBE results than the DFT-LDA ones; nevertheless, all the
zero-point-pressure results are in good agreement.

The DFT quasiharmonic zero-point pressures have been added to the
corresponding static-lattice pressures to give the final EOS's.  The
DFT-PBE quasiharmonic zero-point pressure has been added to the DMC
static-lattice pressure to give the final DMC EOS\@.  For the pair
potentials, the quasiharmonic zero-point pressure calculated using
each pair potential has been added to the corresponding static-lattice
pressure in order to obtain the final EOS.

\section{Zero-temperature EOS of neon \label{sec:neon_EoS}}

\begin{figure}
\begin{center}
\includegraphics*{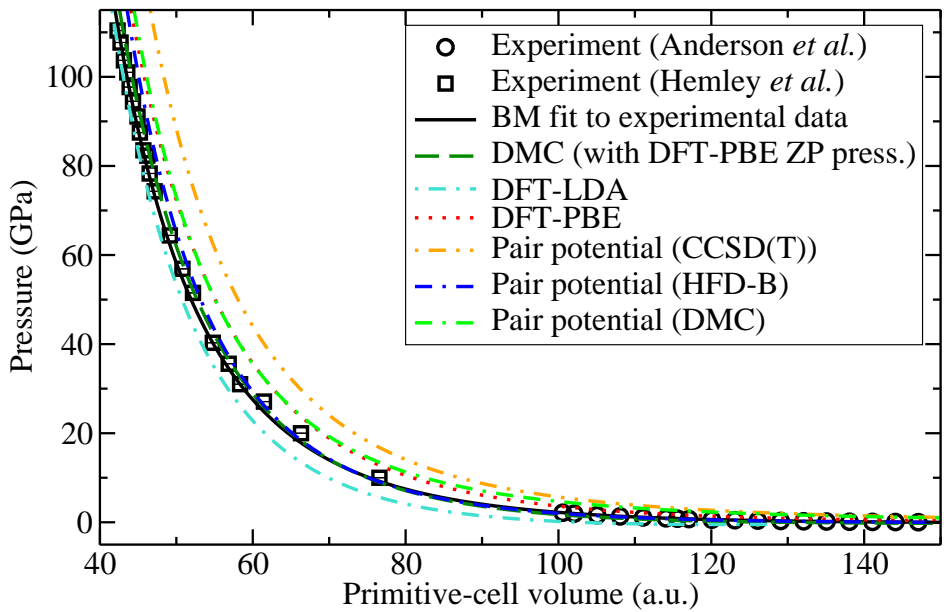}
\caption{(Color online) EOS of FCC neon, obtained by experiment and
  various theoretical techniques.
  \label{fig:neon_p_of_V_inc_ZPP}}
\end{center}
\end{figure}

\begin{figure}
\begin{center}
\includegraphics*{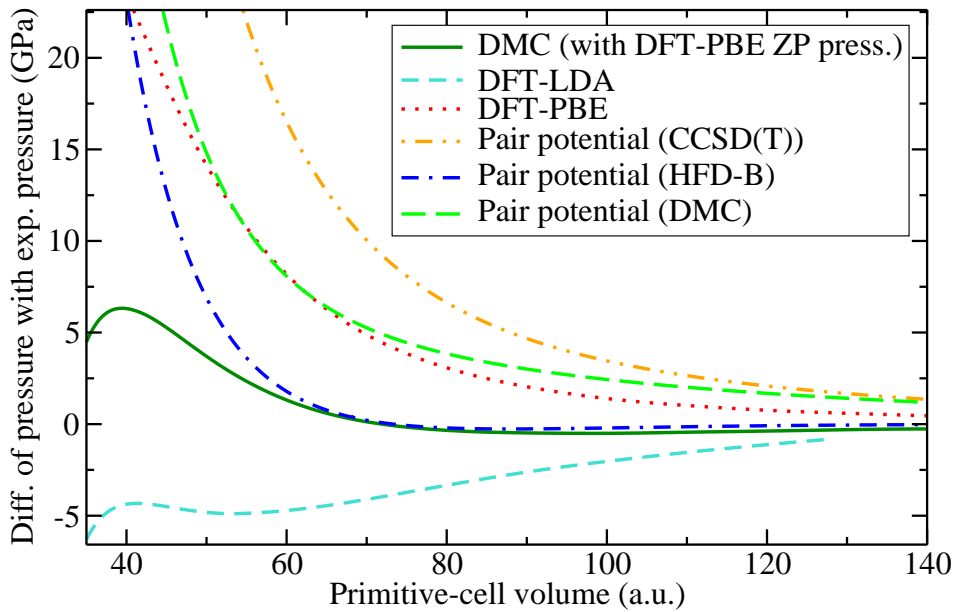}
\caption{(Color online) Comparison of theoretical EOS's of FCC
  neon. The difference of the calculated pressure is plotted against
  the experimental pressure as a function of
  volume. \label{fig:neon_p_of_V_inc_ZPP_new}}
\end{center}
\end{figure}

Zero-temperature EOS's for solid neon, calculated using DFT-LDA,
DFT-PBE, DMC [extrapolated to infinite system size using
Eq.~(\ref{eqn:pressure_extrap})], and pair potentials are shown in
Fig.~\ref{fig:neon_p_of_V_inc_ZPP}, and the differences of the
theoretical EOS's with the experimental EOS are plotted in
Fig.~\ref{fig:neon_p_of_V_inc_ZPP_new}.  (The low-density experimental
pressure-volume data of Anderson \textit{et al.}\cite{anderson}\ shown
in Fig.~\ref{fig:neon_p_of_V_inc_ZPP} were obtained at 4.2~K, while
the high-density experimental data of Hemley \textit{et
al.}\cite{hemley_1989}\ were obtained at 300~K\@.  Hemley \textit{et
al.}\ reduced their pressure-volume data to the zero-temperature
isotherm using a Mie-Gr\"uneisen model and fitted their results to a
third-order Birch-Murnaghan EOS, with the zero-pressure primitive-cell
volume and bulk modulus being the values obtained by Anderson
\textit{et al.}  The resulting EOS\cite{footnote_exp_eos} is valid at
both low and high densities and is regarded as being the definitive
experimental EOS\@.)  The parameter values for a Vinet fit
(Eq.~(\ref{eqn:Vinet_EOS})) to our DMC data (including the DFT-PBE
ZPE) are $V_0=128.52597$~a.u., $B_0=2.7539319$~GPa, and
$B_0^\prime=7.6510744$.

At low densities the DFT-LDA and DFT-PBE EOS's differ markedly.  The
strong dependence of the DFT results on the choice of
exchange-correlation functional implies that the description of van
der Waals bonding within DFT is unreliable, as one would expect, given
the local nature of the approximations to the exchange-correlation
functional.  DMC produces a considerably more accurate EOS than DFT,
suggesting that DMC is capable of giving a proper description of van
der Waals bonding.  The HFD-B pair potential\cite{aziz_1989} gives an
EOS of similar accuracy to the DMC EOS at low to intermediate
densities.  At higher densities, the EOS calculated using DMC is
better than any of the pair-potential EOS's.  Although the difference
between the DMC and experimental pressures is significant at high
densities, it should be emphasized that the fractional error remains
small.

The pair potential calculated using CCSD(T) theory\cite{cybulski}
gives a poorer EOS than the DMC-generated pair potential.  On the
other hand, the EOS obtained using the DMC pair potential is
significantly poorer than the HFD-B EOS.  Taken together with the fact
that the direct DMC EOS is excellent, this suggests that many-body
interactions play a significant role in solid neon, and that such
interactions are included to some extent in the HFD-B potential.

\section{Conclusions \label{sec:conclusions}}

We have performed DMC calculations of the energy of FCC solid neon as
a function of the lattice constant and the energy of the neon dimer as
a function of atomic separation.  Other calculations using DFT methods
and pair potentials have been performed to evaluate the ZPE and for
comparison purposes.

We have calculated the phonon dispersion curves of solid neon using
the DFT-LDA and DFT-PBE methods, the HFD-B pair potential, and
CCSD(T)- and DMC-derived pair potentials.  We believe the results
obtained with the HFD-B pair potential are likely to be the most
accurate.  DFT-PBE gives more accurate dispersion curves than DFT-LDA,
for which the phonon frequencies are too low.  The dispersion curves
obtained with the DMC pair potential are more accurate than those
obtained using either DFT or the CCSD(T) pair potential.   We have
calculated the ZPE of solid neon using the DFT-LDA and DFT-PBE
methods, and the HFD-B, CCSD(T), and DMC pair potentials, within the
quasiharmonic approximation.  At low pressures the ZPE depends on the
calculation method used, but the contribution to the EOS is small,
while at high pressures the dependence on the calculation method is
relatively weak, although the contribution of the ZPE to the EOS is
significant.  The Einstein model gives ZPE's in very good agreement
with the quasiharmonic values over the pressure range considered.

We have calculated the zero-temperature EOS of solid neon using the
DFT and DMC methods, including corrections for the ZPE.  We have shown
that the DFT results depend strongly on the choice of
exchange-correlation functional, while the DMC results are close to
the experimental EOS\@.  We therefore have evidence that DMC gives a
better description of van der Waals bonding in real materials than
DFT\@.  At high pressures the DMC EOS is closer to the experimental
results than the EOS obtained using the HFD-B pair potential. However,
the statistical errors of about 0.0002~a.u.\ in the DMC energy data
for solid neon are too large to determine an accurate value for the
lattice constant of the solid.  We have shown that the neon pair
potential determined by DMC calculations gives a more accurate EOS
than the pair potential determined by CCSD(T) calculations, although
the DMC pair-potential results are not as accurate as those obtained
using the semiempirical HFD-B potential.  Overall, our results
demonstrate the accuracy and reliability of the DMC method and the
high quality of the neon pseudopotentials that we have used.

\section{Acknowledgments}

Financial support has been provided by the Engineering and Physical
Sciences Research Council (EPSRC), UK\@. Computing resources have been
provided by the Cambridge-Cranfield High Performance Computing
Facility. We thank J.~R. Trail for providing the relativistic
Hartree-Fock pseudopotential used in this work.


\begin{thebibliography}{9}
\bibitem{footnote_aziz} The short-range modifications to the HFD-B
potential proposed by Aziz and Slaman are irrelevant in this work,
because the potential is only modified at a separation of less than
2.36~a.u., corresponding to an FCC primitive-cell volume of 9.32~a.u.
\bibitem{footnote_offset} We have determined the constant offset to
the energy data during the fitting procedure instead of subtracting
the energy of two isolated neon atoms in order to improve the accuracy
of the fit.  The fixed-node error in a neon dimer is larger than the
fixed-node error in an isolated neon atom.  Therefore a direct
evaluation of the absolute value of the pair potential is a (slight)
systematic overestimate.
\end{thebibliography}
\end{document}